\documentclass{ws-procs9x6-cpt16}
\usepackage{graphicx}

\begin{document}

\newcommand{\refeq}[1]{(\ref{#1})}
\def\etal {{\it et al.}}

\title{Lorentz-Violating QCD Corrections to Deep Inelastic Scattering\\}

\author{A.R.\ Vieira}

\address{Indiana University Center for Spacetime Symmetries,\\
Bloomington, IN 47405, USA}

\address{Departamento de F\'isica-ICEx, 
Universidade Federal de Minas Gerais,\\
Belo Horizonte, MG 31.270-901, Brasil}

\begin{abstract}
In this work, 
we present CPT- and Lorentz-violating corrections to observable quantities in electron-proton scattering. We
also show how the theoretical prediction can be used together with data to establish bounds on a coefficient for CPT and Lorentz violation in the QCD sector.
\end{abstract}

\phantom{}\vskip10pt\noindent
Unlike the QED sector of the SME, the quark and gluon sectors of the QCD extension are not stringently constrained.\cite{Tables} Most of the coefficients of the QCD 
sector are effective and obtained from composite objects. One reason is that the QCD Hilbert space contains baryons and mesons rather than quarks at low
energies. Thus, as a first step, we can consider CPT and Lorentz violation (LV) in a process where we can access the quark structure of those composite objects,
Deep Inelastic Scattering (DIS). Electron-proton ($e^- P$) scattering, for instance, gives us information about QCD and the quark structure of the proton. It 
is also a high energy process so that we can treat the QCD coupling $g_s$ perturbatively. The zeroth order, $g^0_s$, is the so-called {\it parton} model. 
Considering the QCD extension,\cite{Alan} a Lorentz-volating version of the parton model and its radiative corrections can be obtained from the 
lagrangian
\begin{equation}
\mathcal{L}_{quark}=\tfrac{1}{2}i(g^{\mu\nu}+ c^{\mu\nu}_Q) (\overline{\psi}\gamma_{\mu}\overleftrightarrow{D}_{\nu}\psi+2iQ_f\overline{\psi}\gamma_{\mu}A_{\nu}\psi),
\label{eq1}
\end{equation}
where $D^{\mu}=\partial^{\mu}+\frac{1}{2}ig_s A^{\mu}_i \lambda_i$ is the covariant derivative and $c^{\mu\nu}_Q$ is the CPT-
and Lorentz-violating quark coefficient. In the high-energy limit, the photon energy $Q^2=-q^2\rightarrow  \infty$ and we can neglect $g_s$,
considering that quarks only interact with the photon by means of their charge $Q_f$.

The unpolarized differential cross section of $e^- P$ scattering is
\begin{equation}
\frac{d^2\sigma}{dxdy}=\frac{\alpha^2 y}{(Q^2)^2}L^{\mu\nu}\ {\rm Im}~ W_{\mu\nu},
\label{eq2}
\end{equation}
where $L^{\mu\nu}=2(k^{\mu}k'^{\nu}+k^{\nu}k'^{\mu}-k\cdot k' g^{\mu\nu})$ is the lepton tensor, $y=\frac{P\cdot q}{P\cdot k}$, $x=\frac{-q^2}{2P\cdot q}$
is the Bjorken scale and $W^{\mu\nu}=i\sum_{spins}\int d^4x\ e^{iq\cdot x}\langle P| J^{\mu}(x)J^{\nu}(0)|P \rangle$ is the proton tensor. All the LV
information is in the parton-photon coupling, $J^{\mu}(x)=Q_f\bar{\psi}(x)\Gamma^{\mu}\psi(x)$. The momenta $P$ and $k$ ($k'$) are the proton and electron 
initial (final) energies.

In Eq.\ \eqref{eq2}, we divided by the flux factor $F=2s$. Some care is required in defining $F$, which is modified by LV.\cite{CK} 
However, in the present situation, the SME is being considered in the CPT-even quark sector. The 
DIS process assumes that a short-wavelength photon only sees the quark structure. We do not have to consider LV in $F$ since it is 
defined according to the proton initial state and the whole proton would be perceived only by a long-wavelength photon. Moreover, the proton coefficient 
$c^{\mu\nu}_P$ is well constrained\cite{Tables} and can be neglected compared to the quark one. 

Calculating the explicit form of $W^{\mu\nu}$ is challenging. It represents our ignorance in the photon-proton interaction. As we stated before, 
we make use of perturbation theory, where the parton model is the zeroth order contribution to the process. It allows us to rewrite $W^{\mu\nu}$ as
\begin{equation}
W^{\mu\nu}\approx i \int d^4 x\ e^{iq\cdot x}\int^1_0 d\xi\sum_f \frac{f_f(\xi)}{\xi}\langle q_f(\xi P)|J^{\mu}(x)J^{\nu}(0)|q_f(\xi P)\rangle,
\label{eq3}
\end{equation}
where $f_f(\xi)$ is the parton distribution function (PDF), the probability of finding a parton $f$ carrying a momentum $\xi P$. 

In Eq.\ \eqref{eq3}, there is a sum over flavors. The quark sector of the SME allows a different coefficient for each
flavor.\cite{Alan} In this case, it is impractical to extract the coefficient from the sum over flavors. If we want to consider one coefficient for each quark, up 
and down, we find that the up charge and the two up quarks in the proton make $c^{\mu\nu}_U$ one order of magnitude bigger than $c^{\mu\nu}_D$ and so 
taking only one coefficient is essentially assuming that $c^{\mu\nu}_Q\approx c^{\mu\nu}_U$.

When we take the imaginary part of $W^{\mu\nu}$, we find that the full fermion propagator gives us a delta function corrected by LV, $\delta( -Q^2+2\xi P\cdot 
q+2c^{qq}+2\xi (c^{qP}+c^{Pq})+ 2\xi^2 c^{PP})$. Consequently, the Bjorken scale is also corrected by a factor $x_c=\frac{2}{ys}(x c^{Pq}+x c^{qP}+ c^{qq})$, where
$s=2k\cdot P$ and $c^{\mu\alpha}_Q p_{\alpha}\equiv c^{\mu p}$. Therefore, the LV correction can be seen as a tree-level violation of Bjorken scaling. We can also confirm this after computing the differential cross section in Eq.\ \eqref{eq2}
\begin{align}
&\frac{d^2\sigma}{dxdy}=\frac{\alpha^2 y}{Q^4}{\rm Im}~ W_2\Big[\frac{1}{2}\left(1-\frac{2}{ys}(c^{Pq}+c^{qP}+2x c^{PP})\right)s^2(1+(1-y)^2)
\nonumber\\
&-2xys c^{PP}-2M^2(c^{kk'}+c^{k'k})+\frac{2s}{x}(1-y)c^{kk}+2s(c^{k'P}+c^{Pk'})
\nonumber\\
&-\frac{2s}{x}c^{k'k'}+2s(1-y)(c^{kP}+c^{Pk})\Big]-\frac{\alpha^2 ys^2}{2Q^4}x_c\frac{d{\rm Im}~ W_2}{dx}(1+(1-y)^2),
\label{eq4}
\end{align}
where $W_2$ is one of the proton structure functions. Its derivative comes from the expansion at first order in $c$ of the whole expression, ${\rm Im}~ 
W^c_2=\frac{4\pi}{ys}\sum_f Q^2_f (x-x_c) f_f(x-x_c)$.

We see that Eq.\ \eqref{eq4} is symmetric on the $c$ indices as it should be,
since its antisymmetric part in the lagrangian \eqref{eq1} can be removed by a field 
redefinition. To compare Eq.\ \eqref{eq4} and $W_2$ with data collected at accelerators, we first choose a frame. For 
instance, this can be the proton rest frame for measurements with the single-arm experiment at SLAC. However, for current data on the DIS 
cross section measured at HERA, the proton is not at rest and has opposite momentum to the initial electron momentum. We also must consider the sidereal time 
variation of $c^{\mu\nu}_Q$, which oscillates as the Earth rotates. Therefore, making a transformation between the Earth frame and 
the canonical Sun-centered frame, we can determine how the laboratory components $c^{\mu\nu}_Q$ change with sidereal time.

We can then use the data collected on the $e^-P$ cross section to establish bounds on $c^{\mu\nu}_Q$. As presented above, 
the LV corrections to this cross section
manifest themselves as a violation of Bjorken scaling. At tree level, we can verify that the usual SM results for the reduced cross section and $\nu W_2$
are independent of $Q^2$. The LV correction to these two quantities introduces a nontrivial dependence on $Q^2$. If 
we fit the data on $W_2$ as a straight line, i.e., $\nu W_2(x,Q^2)=a+bQ^2$, the slope $b$ is very small and can be used to constrain $c^{\mu\nu}_Q$. The 
actual analysis\cite{paper} considers a nontrivial and unknown dependence on $Q^2$ and it is used to constrain the components of $c^{\mu\nu}_Q$ with precision 
of $10^{-5}$ to $10^{-7}$.

\end{document}